\documentclass[lettersize,journal]{IEEEtran}

\usepackage{amsmath,amsfonts,amssymb,amsthm}
\usepackage{textcomp}
\usepackage{bm} 
\usepackage{verbatim}

\usepackage{graphicx}
\usepackage{array}
\usepackage{booktabs} 
\usepackage{multirow}
\usepackage{makecell}
\usepackage{stfloats}
\usepackage{float}
\usepackage[caption=false,font=normalsize,labelfont=sf,textfont=sf]{subfig}

\usepackage{algorithm}
\usepackage{algpseudocode}

\usepackage[numbers,sort&compress]{natbib}
\usepackage{url}

\usepackage[colorlinks, linkcolor=black, anchorcolor=black, citecolor=black, urlcolor=blue]{hyperref}

\def\BibTeX{{\rm B\kern-.05em{\sc i\kern-.025em b}\kern-.08em
		T\kern-.1667em\lower.7ex\hbox{E}\kern-.125emX}}

\usepackage{balance} 

\begin{document}


\title{\huge{Multiple-Mode Affine Frequency Division Multiplexing \\with Index Modulation}}
\author{
		Guangyao Liu,
		Tianqi Mao,~\IEEEmembership{Member,~IEEE},
		Yanqun~Tang,\\
		Jingjing Zhao,~\IEEEmembership{Member,~IEEE},
		Zhenyu~Xiao,~\IEEEmembership{Senior Member,~IEEE}

\thanks{\emph{(Corresponding authors: Tianqi Mao, Zhenyu Xiao.)}}
\thanks{G. Liu, J. Zhao and Z. Xiao are with the School of Electronic and Information Engineering and the State Key Laboratory of CNS/ATM, Beihang University, Beijing 100191, China (e-mails: \{liugy,jingjingzhao,xiaozy\}@buaa.edu.cn).}
\thanks{T. Mao is with Greater Bay Area Innovation Research Institute of BIT, Zhuhai 519000, China, and is also with Beijing Institute of Technology (Zhuhai), Zhuhai 519088, China (e-mail: maotq@bit.edu.cn).}
\thanks{Y. Tang is with the School of Electronics and Communication Engineering, Sun Yat-sen University, Shenzhen 518107, China (e-mail: tangyq8@mail.sysu.edu.cn).}

}

\IEEEaftertitletext{\vspace{-2em}}
\maketitle

\begin{abstract}
Affine frequency division multiplexing (AFDM), a promising multicarrier technique utilizing chirp signals, has been envisioned as an effective solution for high-mobility communication scenarios. In this paper, we develop a multiple-mode index modulation scheme tailored for AFDM, termed as MM-AFDM-IM, which aims to further improve the spectral and energy efficiencies of AFDM. Specifically, multiple constellation alphabets are selected for different chirp-based subcarriers (chirps). Aside from classical amplitude/phase modulation, additional information bits can be conveyed by the dynamic patterns of both constellation mode selection and chirp activation, without extra energy consumption. Furthermore, we discuss the mode selection strategy and derive an asymptotically tight upper bound on the bit error rate (BER) of the proposed scheme under maximum-likelihood detection. Simulation results are provided to demonstrate the superior performance of MM-AFDM-IM compared to conventional benchmark schemes.
\end{abstract}


\vspace*{-2mm}
\section{Introduction}
\IEEEPARstart{T}{he} paradigm for sixth-generation (6G) wireless networks is predicated on delivering ubiquitous connectivity for high-mobility applications like vehicle-to-everything (V2X) and high-speed railway communications \cite{9970355}. These scenarios are characterized by significant Doppler shifts that induce time-frequency doubly-dispersive channels, posing a formidable obstacle to reliable data transmission. Whilst orthogonal frequency division multiplexing (OFDM), the cornerstone of 4G and 5G, effectively mitigates the time dispersion effects, its subcarrier orthogonality can be inevitably destroyed by strong Doppler shifts, leading to severe inter-carrier interference (ICI) \cite{10054381}. This fundamental limitation has motivated a paradigm shift in waveform design for 6G communication scenarios. 

Under this background, affine frequency division multiplexing (AFDM) has emerged as a promising candidate technology for doubly-dispersive channels \cite{bemani2021afdm}. By leveraging the unique properties of discrete affine Fourier transform (DAFT), AFDM can separate all channel paths, thereby achieving full channel diversity and demonstrating robust resilience to both time and frequency selective fading. Compared with existing full-diversity counterparts, e.g. orthogonal time-frequency space (OTFS) \cite{bemani2023affine}, AFDM exhibits a compelling performance-to-complexity trade-off. Therefore, there have been increasing interests on AFDM algorithms, including multiple access \cite{10566604}, integrated sensing and communication (ISAC) \cite{10551402,zhang2025afdm}, and channel estimation \cite{10557524,10711268}. Despite these efforts, the potential of AFDM in further enhancing spectral and energy efficiency remains largely untapped, which motivates further innovations on the modulation philosophies. \looseness=-1

The index modulation (IM) techniques, which convey energy-free bits through the activation patterns of transmission resources \cite{bacsar2013orthogonal,10381617}, can achieve significant gains in both spectral and energy efficiencies \cite{8417419,9398861}. Therefore, the IM philosophy has been recently introduced to AFDM. 
In \cite{10342712}, the authors proposed an IM-aided AFDM scheme by embedding the energy-free bits into the activation pattern of AFDM chirps, which achieved BER performance improvement over classical AFDM. \cite{10570960} further proposed a distributed IM scheme by applying distributed mapping for IM-aided AFDM to enhance BER performance. However, unlike recent IM-aided distinguishable modes in OFDM that activate all subcarriers \cite{7547943,7936676,9153161}, the above schemes deactivated some chirps, which caused undesirable squander of the scarce frequency resources. In \cite{anoop2025dual}, the authors utilized the dual-mode IM philosophy for spectral efficiency enhancement, where two distinct constellations are employed within AFDM frame to activate all chirps, demonstrating the potential of IM-aided distinguishable modes in AFDM.  

Against this background, we further extend the IM-aided distinguishable modes philosophy to AFDM by proposing a novel multiple-mode IM transmission strategy, referred to as MM-AFDM-IM. In the proposed scheme, different chirps in AFDM block can select multiple constellation alphabets. Therefore, the information bits can be conveyed not only by the traditional amplitude/phase modulation, but also by the dynamic patterns of both multiple constellation mode selection and chirp activation. Moreover, the mode selection strategy is discussed and the asymptotically tight upper bound on the theoretical BER of the proposed scheme under maximum-likelihood (ML) detection is derived based on pairwise error probability (PEP) analysis. The simulation results confirm that the proposed MM-AFDM-IM scheme achieves superior performance compared to conventional benchmark schemes under doubly-dispersive channels.     


\vspace*{-2mm}
\section{Proposed MM-AFDM-IM Scheme}

\subsection{Transmitter}

\begin{figure*}[!t]   
\vspace*{-4mm}
\centering
\includegraphics[width=7in]{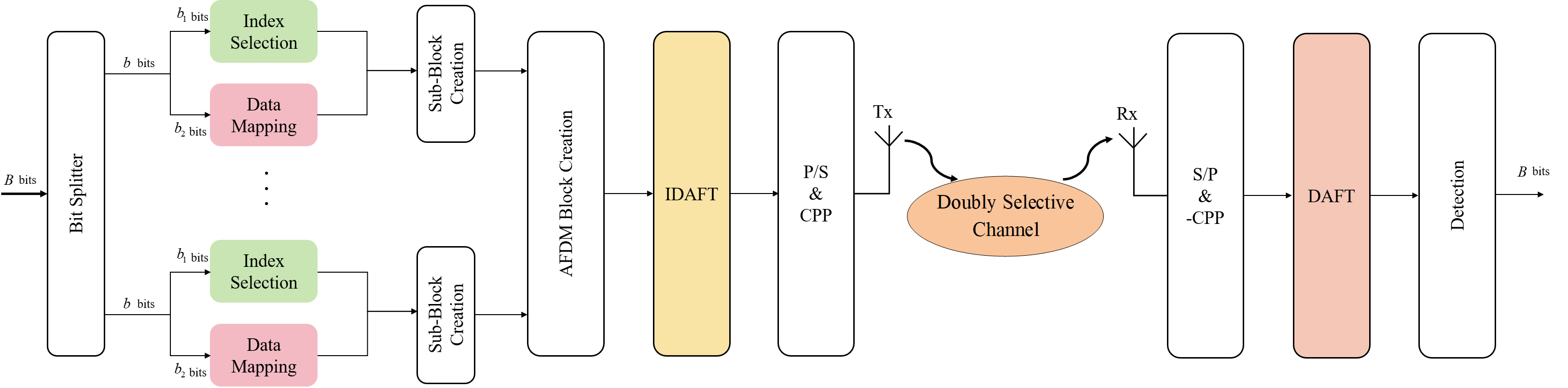}
\caption{Transmitter structure of the proposed MM-AFDM-IM scheme.}
\label{Transmitter}
\vspace*{-3mm}
\end{figure*}	

Consider an AFDM frame in the DAF domain, which consists of $N$ chirps for transmitting totally $B$ information bits. These $N$ chirps are equally partitioned into $G$ sub-blocks, each comprising $n = N/G$ chirps. As seen in Fig.~\ref{Transmitter}, the $B$ information bits for each AFDM frame are first divided into $G$ sub-streams, each corresponding to one sub-block and containing $b = B/G$ bits. Afterwards, each $b$-bit sub-stream is partitioned into $b_1$ index bits and $b_2$ symbol bits, such that $b = b_1 + b_2$. 

Unlike classical AFDM with fixed constellations, the proposed MM-AFDM-IM employs $M$ $U$-ary constellation modes $\mathcal{M}_1, \mathcal{M}_2, \ldots, \mathcal{M}_M$, where $\mathcal{M}_1 \cap \mathcal{M}_2 \cap \ldots \cap \mathcal{M}_M = \varnothing$. For each sub-block, $k$ out of $M$ available modes are selected for different subcarriers according to the index bits. Specifically, the $b_1$ bits determine both the mode activation pattern (MAP) and chirp arrangement pattern (CAP) of each sub-block, where MAP represents which $k$ modes are activated for each transmission, and CAP denotes the arrangement of the $k$ activated modes assigned to the $n$ chirps. For simplicity, we assume that $n$ is an integer multiple of $k$, and each constellation mode is allocated to $n/k$ chirps.
Then the numbers of possible MAP and CAP realizations can be respectively calculated as
\begin{equation}\label{N_MAP} 
	N_\text{MAP} = \binom{M}{k}
\end{equation} 
and
\begin{align}\label{N_CAP} 
&N_\text{CAP} = \binom{n}{n/k} \binom{n - n/k}{n/k} \cdots \binom{n/k}{n/k} \notag \\
&= \frac{n!}{(n/k)! (n - n/k)!} \cdot \frac{(n - n/k)!}{(n/k)! (n - 2n/k)!} \frac{(n/k)!}{(n/k)! \cdot 0!} \notag \\
&= \frac{n!}{((n/k)!)^k} .
\end{align}
By incorporating (1) and (2), $b_1$ can be expressed as
\begin{equation}\label{b1} 
	{b_1} = \left\lfloor {{{\log }_2}\left( {\binom{M}{k} \frac{{n!}}{{{{((n/k)!)}^k}}}} \right)} \right\rfloor.
\end{equation}  

In the transmitter, the $b_1$ bits related to the $g$-th ($g \in \{1,\ldots,G\}$) sub-block are mapped to a specific MAP and CAP combination using a joint selection mechanism, which is summarized below: 

1) The $b_1$ bits are converted to a decimal number $d_g$; 

2) Find the unique index pair  $({I}_{\text{MAP},g},{I}_{\text{CAP},g})$ of the selected MAP and CAP that satisfies 
\begin{equation}\label{joint_decomp} 
d_g = {I}_{\text{MAP},g} + N_\text{MAP} {I}_{\text{CAP},g},
\end{equation}
where ${I}_{\text{MAP},g} \in \{0, \ldots, N_\text{MAP} - 1\}$ and ${I}_{\text{CAP},g} \in \{0, \ldots, N_\text{CAP} - 1\}$ denote the indices in the feasible set of MAP and CAP, respectively; 

3) Determine the MAP and CAP according to the look-up table, which contains the feasible set of MAP and CAP. 
An example of look-up table for $M = 4, n = 4, k = 2, U = 2$ is presented in Table~\ref{Tab1}, where $S_1^{(j)},S_2^{(j)},\ldots,S_k^{(j)}$ represent the $j$-th symbol of $k$ selected modes with $j \in \{1, \ldots, n/k\}$. 
\begin{table}[t]
\centering
\caption{A Joint Selection Example of MAP and CAP for $M = 4, n = 4, k = 2, U = 2$.}
\label{Tab1}
\begin{tabular}{cccc}
\toprule
${I}_{\text{MAP},g}$ & $\text{MAP}$ & ${I}_{\text{CAP},g}$ & $\text{CAP}$ \\
\midrule
0 & [$\mathcal{M}_1$, $\mathcal{M}_2$] & 0 & [$S_1^{(1)}$, $S_1^{(2)}$,$S_2^{(1)}$,$S_2^{(2)}$] \\
1 & [$\mathcal{M}_1$, $\mathcal{M}_3$] & 1 & [$S_1^{(1)}$, $S_2^{(1)}$,$S_1^{(2)}$,$S_2^{(2)}$] \\
2 & [$\mathcal{M}_1$, $\mathcal{M}_4$] & 2 & [$S_1^{(1)}$, $S_2^{(1)}$,$S_2^{(2)}$,$S_1^{(2)}$] \\
3 & [$\mathcal{M}_2$, $\mathcal{M}_3$] & 3 & [$S_2^{(1)}$, $S_1^{(1)}$,$S_1^{(2)}$,$S_2^{(2)}$] \\
4 & [$\mathcal{M}_2$, $\mathcal{M}_4$] & 4 & [$S_2^{(1)}$, $S_1^{(1)}$,$S_2^{(2)}$,$S_1^{(2)}$] \\
5 & [$\mathcal{M}_3$, $\mathcal{M}_4$] & 5 & [$S_2^{(1)}$, $S_2^{(2)}$,$S_1^{(1)}$,$S_1^{(2)}$] \\
\bottomrule
\end{tabular}
\end{table}

Then, the ${b_2} = n{\log _2}U$ bits are modulated onto the chirps according to the selected MAP and CAP, which yields the $g$-th AFDM sub-block, expressed as
\begin{equation}\label{subblock} 
{{\mathbf{x}}_g } = {[{x_g }(1), \ldots ,{x_g }(n)]^\mathrm{T}}, 
\end{equation}
where \({x_g }(i) \in {\mathcal{M}_{{i_g }}}(i = 1,2, \ldots ,n)\) represents the data symbol of the $i_g$-th mode, and $i_g \in \{1,2,\ldots,M\}$ denotes the index of selected constellation mode for the $i$-th chirp within the $g$-th sub-block.

The generated $G$ sub-blocks are then assembled into an $N \times 1$ transmission vector in the discrete affine Fourier domain, expressed as
\begin{align} 
\mathbf{x} = 
\left[ \mathbf{x}_1^\mathrm{T},\mathbf{x}_2^\mathrm{T}, \ldots, \mathbf{x}_G^\mathrm{T} \right]^\mathrm{T} 
= \left[ x(1), \ldots, x(N) \right]^\mathrm{T}.
\end{align}

The time-domain transmitted signal is then obtained by performing an $N$-point inverse-DAFT (IDAFT) on $\mathbf{x}$, formulated as
\begin{equation}\label{IDAFT} 
	s[n] = \frac{1}{{\sqrt N }}\sum\limits_{m = 0}^{N - 1} x [m]{e^{\textsf{j} 2\pi \left( {{c_1}{n^2} + {c_2}{m^2} + \frac{nm}{N}} \right)}},
\end{equation}
where $n \in \{0, 1, \ldots, N-1\}$, $c_1$ and $c_2$ denote the post- and pre-chirp parameters of the DAFT, respectively. 

The matrix representation of (\ref{IDAFT}) is given by 
\begin{equation} 
	\mathbf{s} = \mathbf{\Lambda}_{c_1}^\mathrm{H} \mathbf{F}^\mathrm{H} \bm{\Lambda}_{c_2}^\mathrm{H} \mathbf{x} = \mathbf{A}^\mathrm{H} \mathbf{x},
\end{equation}
where $\mathbf{F}$ denotes the discrete Fourier transform (DFT) matrix with entries $F(m,n) = e^{-\textsf{j} 2 \pi m n/N}/\sqrt{N}$, $\bm\Lambda_{c_i} = \operatorname{diag}(e^{-\textsf{j} 2\pi c_i n^2}, n = 0, \ldots, N-1)$ is the post- and pre-chirp diagonal matrices for $i=1$ and $i=2$, respectively. Finally,  $\mathbf{A} = \bm{\Lambda}_{c_2} \mathbf{F} \mathbf{\Lambda}_{c_1}$ represents the DAFT operator in matrix form.

Before transmission of time-domain signals, the proposed MM-AFDM-IM scheme needs to append a chirp-periodic prefix (CPP) to effectively mitigate the impact of multipath propagation, whose length is assumed to exceed the maximum delay spread of the channel.

\subsection{Channel and Receiver}
In high-mobility scenarios, the transmitted typically experience a time-frequency doubly dispersive channel due to pronounced Doppler shifts and multipath propagation. Consider a doubly dispersive channel consisting of $P$ propagation paths, each characterized by a triplet $\{h_p, \tau_p, \nu_p\}$ representing the complex channel coefficient, delay, and Doppler shift, respectively. Then the mathematical channel model is given by
\begin{equation}\label{eqCIR}  
	h(\tau, \nu) = \sum_{p=1}^P h_p \delta(\tau-\tau_p) \delta(\nu-\nu_p),
\end{equation}
where $\delta\{\cdot\}$ denotes the Dirac delta function, and $h_p\! \sim\! \mathcal{CN}(0,1/P)$. We normalize the delay $\tau_p$ and Doppler shift $\nu_p$ of each path as $d_p = \tau_p \Delta f$ and $\alpha_p = N T \nu_p$, respectively, where $\Delta f$ is the subcarrier spacing, and $T$ is the sampling interval. The normalized Doppler shift $\alpha_p$ is modeled as the sum of an integer component $l_p$ and a fractional component $\varepsilon_p$, i.e., $\alpha_p = l_p + \varepsilon_p$. Besides, $d_p$ and $\alpha_p$ are constrained by $d_p \in [0, d_{\max}]$ and $\alpha_p \in [-\alpha_{\max}, \alpha_{\max}]$, given that $d_{\max}$ and $\alpha_{\max}$ denoting the maximum normalized delay and Doppler shift, respectively.

Based on the aforementioned model, the received signals in time domain can be expressed as 
\begin{equation}\label{time-signals} 
	r[n] = \sum\limits_{p = 1}^P {{h_p}} s[n - {d_p}]{e^{ - \textsf{j}2\pi {\nu _p}n/N}} + {w_r}[n],
\end{equation}
where $w_r[{n}]\! \sim\! \mathcal{C N}\left(0, N_0\right)$ represents the noise component. The matrix equivalent of (\ref{time-signals}) is derived as 
\begin{equation}\label{eqMRxS} 
	\mathbf{r} = \sum_{p=1}^{P} h_{p} \mathbf{\Gamma}_{\textnormal{CPP}_{p}} \bm{\Delta}_{\nu_{p}} \mathbf{\Pi}^{d_{p}} \mathbf{s} + \mathbf{w}_r,
\end{equation}  
where $\mathbf{\Gamma}_{\text{CPP}_p}$ is the $N \times N$ diagonal CPP matrix with its $n$-th diagonal element given by
\begin{equation}\label{CPP_matrix} 
	\omega_{p,n} = \begin{cases}e^{-\textsf{j} 2\pi c_1(N^2-2N(d_p-n))},& n<d_p, \\ \qquad \qquad 1, & n\ge d_p . \end{cases}
\end{equation}
Besides, $\bm{\Delta}_{\nu_p} = \operatorname{diag}\left(1, e^{-\textsf{j} 2 \pi \nu_p/N}, \ldots, e^{-\textsf{j} 2 \pi (N-1)\nu_p/N} \right)$ is the Doppler shift matrix, $\mathbf{\Pi}$ denotes a $N \times N$ forward cyclic-shift matrix, and $\mathbf{w}_r$ is the $N\times 1$ noise vector.

By applying an $N$-point DAFT to the received signals, the corresponding DAF-domain representation is obtained as
\begin{align}\label{eqRxS} 
	y[l] = \frac{1}{\sqrt{N}}\sum_{n=0}^{N-1} r[n] e^{-\textsf{j} 2 \pi\left(c_1 n^2+c_2 l^2+ \frac{n l}{N}\right)},
\end{align}
where $l =0, 1, \ldots, N-1$. Then the input-output relationship for the proposed MM-AFDM-IM scheme can be finally formulated as
\begin{align}\label{y_matrix} 
	\mathbf{y} = \mathbf{A} \mathbf{r} = \mathbf{H}_\textnormal{eff}\mathbf{x} + \mathbf{w},
\end{align}
where $\mathbf{y}$ is the $N \times 1$ received signal in DAF domain, $\mathbf{H}_\textnormal{eff} = \sum_{p=1}^P h_p \mathbf{A}\mathbf{\Gamma}_{\textnormal{CPP}_p} \mathbf{\Delta}_{\nu_p} \mathbf{\Pi}^{d_p}\mathbf{A}^\mathrm{H}$ and $\mathbf{w} = \mathbf{A}\mathbf{w}_r$ represents the effective channel matrix and noise in the DAF domain, respectively.

Based on (13), we employ a joint ML detector by searching all the realizations of MAPs, CAPs and data symbols, which can be formulated as
\begin{equation}\label{ML1} 
	\left(\hat{\mathbf{x}},\hat{\mathbf{I}}_{\text{MAP}},\hat{\mathbf{I}}_{\text{CAP}} \right)= \arg \min\limits_{\forall \mathbf{x},\mathbf{I}_{\text{MAP}},\mathbf{I}_{\text{CAP}}}\left\|\mathbf{y}-\mathbf{\hat{H}}_\mathrm{eff} \mathbf{x}\right\|^2 ,
\end{equation}
where ${\mathbf{I}}_{\text{MAP}} = [{I}_{\text{MAP},1},{I}_{\text{MAP},2},\ldots, {I}_{\text{MAP},G}]$ and ${\mathbf{I}}_{\text{CAP}} = [{I}_{\text{CAP},1},{I}_{\text{CAP},2},\ldots,{I}_{\text{CAP},G}]$ represents the index patterns of an MM-OFDM-IM frame. Finally, the totally $B$ information bits can be readily reconstructed by demapping of the detected results.
\vspace*{-1mm}
\section{Performance Analysis}
\subsection{Mode Selection Strategy}
The design of multiple-mode constellations is critical to the performance of the proposed MM-AFDM-IM scheme. Following the principles established in \cite{7936676}, the design objective is to enlarge both the minimum intra-mode Euclidean distance (MIAD) and the minimum inter-mode Euclidean distance (MIRD).

\begin{figure}[!t]
\centering
\includegraphics[width=3.3in]{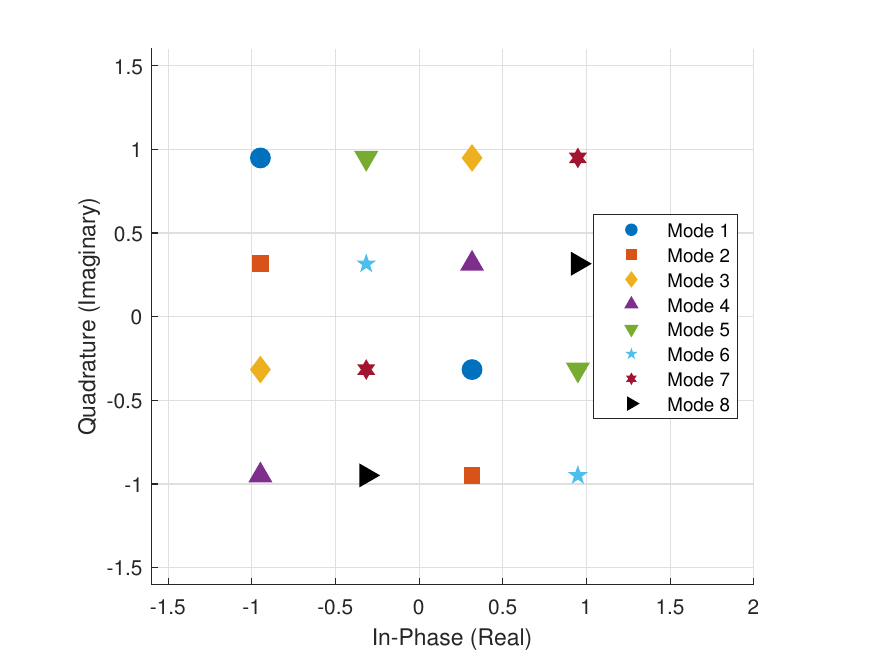}
\vspace*{-3mm}
\caption{An example for 8-mode constellation, which constitutes a 16-QAM format.}
\label{constellation}	
\vspace*{-4mm}
\end{figure}

A practical strategy for generating the distinguishable modes is to partition the standard phase-shift keying (PSK) or quadrature amplitude modulation (QAM) constellations. For the different modes originated from a PSK parent, the MIAD and MIRD can be expressed as
\begin{equation} 
	d_{\text{MIRD}}^{\text{PSK}}(M,U) = d_{\text{MIAD}}^{\text{PSK}}(MU) =  2\sin\left(\frac{\pi}{MU}\right).
\end{equation}
On the other hand, for modes derived from a rectangular QAM constellation, the MIRD and MIAD are calculated as
\begin{equation} 
	d_{\text{MIRD}}^{\text{QAM}}(M,U) = \frac{2 \sqrt{6}}{\sqrt{5MU-4}},
\end{equation}
\begin{equation} 
d^{\text{QAM}}_{\text{MIAD}}(M,U) = 
\begin{cases}
d^{\text{QAM}}_{\text{MIRD}}(M,U) \, \frac{\sqrt{5U}}{2}, & U=2, \\
d^{\text{QAM}}_{\text{MIRD}}(M,U) \, \sqrt{U}, & \text{otherwise}.
\end{cases}
\end{equation}
It can be observed from (15-17) that the QAM-based partitioning yields a larger MIRD for most practical cases (i.e., $MU>4$) and a superior MIAD for $M>4$. Therefore, the QAM-based strategy is mainly considered in this paper for better expected performance. An example for $M=8$ and $U=2$, i.e., 16-QAM-based 8-mode constellations is shown in Fig.~\ref{constellation}.

\vspace*{-3mm}
\subsection{Error Performance Analysis} \label{S3.2}

To evaluate the performance of the proposed MM-AFDM-IM scheme, this section derives an analytical upper bound on the system average bit error probability (ABEP), where the optimal ML detection is considered.

According to (\ref{y_matrix}), the DAF-domain received signals can be re-formulated as
\begin{equation}\label{inputoutput} 
	\mathbf{y} = \sum_{p=1}^P h_p \mathbf{H}_{p}\mathbf{x} + \mathbf{w} =\mathbf{\Phi}(\mathbf{x})\mathbf{h} + \mathbf{w},
\end{equation}
where $\mathbf{H}_{p} = \mathbf{A}\mathbf{\Gamma}_{\textnormal{CPP}_p} \mathbf{\Delta}_{\nu_p} \mathbf{\Pi}^{d_p}\mathbf{A}^\mathrm{H}$ denotes the effective channel of the $p$-th path in the DAF domain, $\mathbf{\Phi}(\mathbf{x}) \in \mathbb{C}^{N \times P}$ is the concatenated matrix constructed as  $\mathbf{\Phi}(\mathbf{x})=\! [\mathbf{H}_{1}\mathbf{x},\mathbf{H}_{2}\mathbf{x}, \ldots, \mathbf{H}_{P}\mathbf{x}] $, and $\mathbf{h}\! =\! [h_{1},h_{2},\ldots,h_{P}]^{\rm T}\! \in\! \mathbb{C}^{P \times 1}$ represents the channel coefficient vector for different paths. 

Based on the expression in (18), the theoretical performance analysis is conducted, where we assume perfect knowledge of the channel state information (CSI) at the receiver, and the average energy of the transmitted vector $\mathbf{x}$ is normalized to unity. A pairwise error event is defined as that $\mathbf{x}$ is incorrectly detected as another transmitted vector realization $\hat{\mathbf{x}}$, caused by misdetections of either the index patterns or the modulated symbols. Correspondingly, the conditional PEP (CPEP) is derived as
\vspace*{-1mm}
\begin{equation}\label{eqCPEP1} 
	\Pr\big(\mathbf{x} \to \hat{\mathbf{x}} | \mathbf{h}\big) = Q\biggl(\sqrt{\frac{\left\|\big(\mathbf{\Phi} (\hat{\mathbf{x}})\! -\! \mathbf{\Phi}(\mathbf{x})\big)\mathbf{h}\right\|^2}{2N_{0}}}\biggr),
\end{equation}
where $Q(\cdot)$ denotes the $Q$-function. A tight and widely-used approximation for $Q(x)$ is given by
\begin{equation}\label{eqQF} 
	Q(x) \approx \frac{1}{12}e^{-x^2/2}+\frac{1}{4}e^{-2x^2/3}.
\end{equation}
Substituting (\ref{eqQF}) into (\ref{eqCPEP1}), the CPEP is approximated as
\begin{equation}\label{eqCPEP2} 
	\Pr\big(\mathbf{x} \to \hat{\mathbf{x}} | \mathbf{h}\big) \approx \frac1{12}e^{-\lambda_1\delta} + \frac14e^{-\lambda_2\delta},
\end{equation}
where $\delta = \left\|\big(\mathbf{\Phi} (\hat{\mathbf{x}})\! -\! \mathbf{\Phi}(\mathbf{x})\big)\mathbf{h}\right\|^2$, $\lambda_1\! =\! 1/({4 N_0})$, and $\lambda_2\! =\! 1/({3 N_0})$. Afterwards, the unconditional PEP (UPEP) can be obtained by averaging the CPEP with the probability density function of $\delta$, $f_{\delta}(\delta)$, which can be formulated as
\vspace*{-1mm}
\begin{align}\label{PEP1} 
	\Pr\big(\mathbf{x} \to \hat{\mathbf{x}}\big) &=\mathrm{E}_\mathbf{h}\Big(\mathbf{x} \to \hat{\mathbf{x}} | \mathbf{h}\Big) \nonumber \\
	&\approx \int_{0}^{+\infty}\biggl(\frac{1}{12}e^{-\lambda_1\delta}+\frac{1}{4}e^{-\lambda_2\delta}\biggr)f_{\delta}(\delta) \mathrm{d} \delta .
\end{align}
By leveraging the definition of the moment-generating function (MGF) $M_{\eta}(s) = \mathrm{E}[e^{s\eta}]= \int_{-\infty}^{+\infty}e^{s\eta}f_{\eta}(\eta )\mathrm{d} \eta$, (\ref{PEP1}) can be further calculated as
\begin{equation}\label{PEP2} 
	\Pr\big(\mathbf{x} \to \hat{\mathbf{x}}\big) \approx \frac{1}{12}M_\delta(-\lambda_1)+\frac{1}{4}M_\delta(-\lambda_2).
\end{equation}

According to \cite[Theorem 2]{10975107}, for a Hermitian matrix $\mathbf{\Theta}$ and a zero-mean complex Gaussian vector $\mathbf{v}$ with a covariance matrix $(1/P)\mathbf{I}_P$, the MGF of the quadratic form $f\! =\! \mathbf{v}^\mathrm{H} \mathbf{\Theta} \mathbf{v}$ has a solution expressed as 
\vspace*{-1mm}
\begin{equation}\label{MGF} 
M_f(s) = \prod_{i=1}^{\gamma} \frac{1}{1 - (s/P)\varepsilon_i},
\end{equation}
where $\varepsilon_i$ and $\gamma$ represent the nonzero eigenvalues and rank of $\mathbf{\Theta}$, respectively.
Note that $\delta$ is a quadratic form of the channel vector, expressed as $\delta = \mathbf{h}^\mathrm{H} \mathbf{\Upsilon } \mathbf{h}$, where $\mathbf{\Upsilon } = \big(\mathbf{\Phi} (\hat{\mathbf{x}})\! -\! \mathbf{\Phi}(\mathbf{x})\big)^\mathrm{H} \big(\mathbf{\Phi} (\hat{\mathbf{x}})\! -\! \mathbf{\Phi}(\mathbf{x})\big)$ is a  Hermitian matrix.

Therefore, by substituting (\ref{MGF}) into (\ref{PEP2}), the UPEP can be finally derived as
\vspace*{-1mm}
\begin{align}\label{PEP_final} 
	& \Pr\big(\mathbf{x} \to \hat{\mathbf{x}}\big) \approx\! \frac{1}{12}\prod_{i=1}^{\gamma} \frac{1}{1+\frac{\lambda_1 \varepsilon_i}{P}}\! +\! \frac{1}{4}\prod_{i=1}^{\gamma} \frac{1}{1\! +\! \frac{\lambda_2 \varepsilon_i}{P}}.
\end{align}
By considering the UPEPs of all possible pairwise error events, the upper bound on the ABEP is derived as
\vspace*{-1mm}
\begin{equation}\label{ABEP} 
	\mathrm{Pr}_{\mathrm{ABEP}} \leq \frac{1}{2^B B} 
	\sum_{\mathbf{x}} \sum_{\hat{\mathbf{x}}} 
	\mathrm{Pr}(\mathbf{x} \to \hat{\mathbf{x}}) \, e(\mathbf{x} \to \hat{\mathbf{x}}),
\end{equation}
where $e(\mathbf{x} \to \hat{\mathbf{x}})$ denotes the number of erroneous bits by wrongly detecting $\mathbf{x}$ as $\hat{\mathbf{x}}$.
\vspace*{-2mm}
\section{Simulation Results}
In this section, we evaluate the BER performance of the proposed MM-AFDM-IM system via Monte Carlo simulations. For performance comparison, several benchmark schemes are considered, including AFDM-IM \cite{10342712}, AFDM-IM-distributed \cite{10570960}, and SuM-OFDM-IM \cite{9153161}. The ML detector is employed for all systems. The carrier frequency is 8 GHz with a chirp subcarrier spacing of 15 kHz. The maximum normalized delay and Doppler shift are set to $d_{\max}=1$ and $\alpha_{\max}=1$, respectively. We set that $c_1 = (2(\alpha_{\max}+1)+1)/2N$ and $c_2$ is an arbitrary irrational number. And the Doppler shift for the $p$-th path is generated according to the Jakes Doppler model as $\alpha_{p} = \alpha_{\max} \cos(\theta_{p})$ with $\theta_{p} \in [-\pi, \pi]$. The QAM-based strategy is employed for MM-AFDM-IM.

\begin{figure}[!t]
\centering
\includegraphics[width=3in]{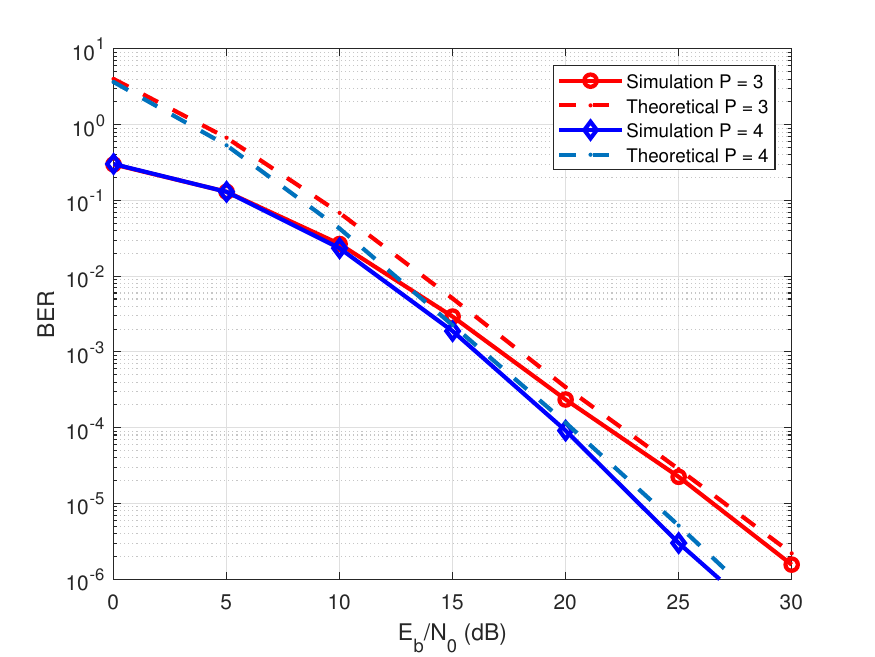}
\vspace*{-4mm}
\caption{Comparison of the theoretical ABEP upper bound and the simulated BER of the proposed MM-AFDM-IM.}
\label{vsTheo.}	
\vspace*{-4mm}
\end{figure}

Fig.~\ref{vsTheo.} validates the derived theoretical upper bound by comparing it with the simulated BER performance of the proposed MM-AFDM-IM scheme for channels with $P=3$ and $P=4$ paths. The system parameters are configured as $(N,M,n,G,k,U) = (4, 4,4,1,2,2)$. In the low-SNR regime, a gap between the analytical bound and the simulation results is observed, which is attributable to the nature of the union bound approximation, known to be less tight at low SNRs. In the high-SNR regime, the theoretical curve serves as a tight upper bound for the simulated performance, confirming the validity of the analysis presented in Subsection~\ref{S3.2}. The results also reveal a notable performance enhancement as the number of paths increases from 3 to 4, stemming from the larger diversity gain achieved with more independent paths.

\begin{figure}[!t]
\centering
\includegraphics[width=3in]{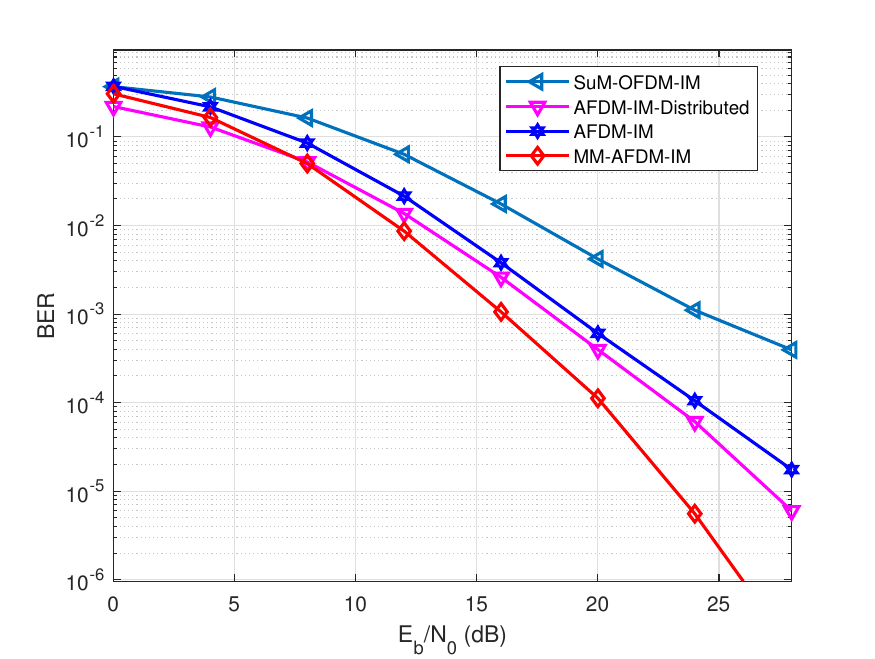}
\vspace*{-4mm}
\caption{Performance comparison between the proposed MM-AFDM-IM and benchmarks using ML detector.}
\label{vsBenchmark}
\vspace*{-5mm}
\end{figure}

In Fig.~\ref{vsBenchmark}, the BER performance of the proposed MM-AFDM-IM is evaluated against AFDM-IM \cite{10342712}, AFDM-IM-distributed \cite{10570960}, and SuM-OFDM-IM \cite{9153161} under a doubly dispersive channel with $P=3$ paths. All systems are configured for a fair comparison at an identical spectral efficiency of 2.25 bit/s/Hz.
The parameter set for the proposed MM-AFDM-IM scheme is $(N,M,n,G,k,U) = (8, 4, 4, 2, 2, 2)$. For the AFDM-IM and AFDM-IM-distributed schemes, 4 out of 8 chirps are activated per block with 8-QAM modulation. For the SuM-OFDM-IM scheme, a 4-mode 16-QAM system with 4 constellation points for each mode is employed.
The results from Fig.~\ref{vsBenchmark} demonstrate the superiority of our proposed scheme. Specifically, MM-AFDM-IM exhibits an SNR gain of more than 1.5 dB at the BER level of $10^{-3}$ relative to the benchmarks. The underlying reason for this enhancement is the scheme's design, which allocates a higher proportion of bits to the index domain. By shifting the information load from the vulnerable constellation symbols to the index bits, the system is less impacted by the impairments caused by delay and Doppler spreads. 

\section{Conclusion}
In this paper, we proposed a novel MM-AFDM-IM scheme to enhance both spectral efficiency and reliability in doubly-dispersive channels. Our proposed scheme utilizes all available chirps for data transmission by embedding information in the joint selection of constellation modes and their permutation patterns, which overcomes the spectral inefficiency of sparse IM approaches. Furthermore, a tight closed-form upper bound on the ABEP was derived based on PEP analysis and verified through simulations. The simulation results have shown that the proposed MM-AFDM-IM exhibits superior error performance in comparison to traditional benchmarks.


\footnotesize
\bibliographystyle{IEEEtran}   
\bibliography{MS-AFDM-IM}   

\end{document}